\documentclass[conference]{IEEEtran}
\IEEEoverridecommandlockouts
\usepackage{cite}
\usepackage{amsmath,amssymb,amsfonts}
\usepackage{algorithmic}
\usepackage{graphicx}
\usepackage{textcomp}
\usepackage{xcolor}
\def\BibTeX{{\rm B\kern-.05em{\sc i\kern-.025em b}\kern-.08em
    T\kern-.1667em\lower.7ex\hbox{E}\kern-.125emX}}
\begin{document}

\title{A tale of two metrics:Polling and financial contributions as a measure of performance\\ 
\textcolor{blue}{\footnotesize \textsuperscript{}Note: This paper has been accepted for publication in 2021 IEEE International Systems Conference (SysCon) proceedings}
\thanks{}
}

\author{\IEEEauthorblockN{Moeen Mostafavi}
\IEEEauthorblockA{\textit{Engineering Systems and Environment } \\
\textit{University of Virginia}\\
Charlottesville, United States \\
mm4ff@virginia.edu}
\and
\IEEEauthorblockN{Maria Phillips}
\IEEEauthorblockA{\textit{Engineering Systems and Environment } \\
\textit{University of Virginia}\\
Charlottesville, United States \\
mp6kv@virginia.edu}
\and
\IEEEauthorblockN{Yichen Jiang}
\IEEEauthorblockA{\textit{Engineering Systems and Environment } \\
\textit{University of Virginia}\\
Charlottesville, United States \\
yj3us@virginia.edu}
\and
\IEEEauthorblockN{Michael D. Porter}
\IEEEauthorblockA{\textit{Engineering Systems and Environment} \\
\IEEEauthorblockA{\textit{School of Data Science }} 
\textit{University of Virginia}\\
Charlottesville, United States \\
mdp2u@virginia.edu}

\and
\IEEEauthorblockN{Paul Freedman}
\IEEEauthorblockA{\textit{Department of Politics } \\
\textit{University of Virginia}\\
Charlottesville, United States \\
freedman@virginia.edu}

}

\maketitle

\begin{abstract}
Campaign analysis is an integral part of American democracy and has many complexities in its dynamics. Experts have long sought to understand these dynamics and evaluate campaign performance using a variety of techniques. We explore campaign financing and standing in the polls as two components of campaign performance in the context of the 2020 Democratic primaries. We show where these measures exhibit represent similar dynamics and where they differ. 
We focus on identifying change points in the trend for all candidates using joinpoint regression models. We find how these change points identify major events such as failure or success in a debate. Joinpoint regression reveals who the voters support when they stop supporting a specific candidate.
This study demonstrates the value of joinpoint regression in political campaign analysis and it represents a crossover of this technique into the political domain building a foundation for continued exploration and use of this method. 
\end{abstract}


\section{Introduction}
The analysis of electoral campaigns has long been a central project for social scientists from a range of disciplines. In analyzing campaigns, a wide range of variables has been brought to bear on measuring candidate performance, including media coverage, social media interactions, fundraising, sanding in the pre-election polls, and more \cite{bartels1988presidential,brown2013does,lipsitz2011filled,jacobson2019politics}. This study explores two of the components of campaign performance in the context of the 2020 Democratic primaries. We examine the timing of increased campaign donations and changes in poll position, asking how they are related? Can one be used to predict changes in the other? Does the relationship between polling and financing look the same at different points in the campaigns or even across different candidates?

Within political science, the literature on campaign finance has focused on the relationship between campaign spending and electoral outcomes (for example, share of vote earned). While fundraising advantages are, not surprisingly, correlated with election outcomes  \cite{bonica2017professional}, this relationship is confounded by the fact that candidate quality (including prior electoral experience and access to party fundraising infrastructure) is correlated with both fundraising ability and ultimate vote share \cite{koerth2018money,jacobson2019politics}. The present study shifts the focus from election outcomes to poll position throughout the primary campaign. The motivation behind this project is to investigate the relationship between alternative measures of campaign performance.

Polls are the main source of evaluation for campaign performance during an election season. They are, of course, an imperfect measure. One challenge in any pre-election poll involves selecting a sample of “likely” voters. Estimating voter likelihood is particularly challenging during primary elections, is usually opaque, and generally varies across pollsters. Another issue involves the phenomenon of “herding,” or the tendency of some polls to down-weight or otherwise adjust findings which diverge from current trends as reported by polling aggregators such as FiveThirtyEight or RealClearPolitics \cite{silver2014fivethirtyeight}. It is useful, therefore, to examine alternative measures of candidate performance. Here we look at campaign finance. To what extent is a candidate’s standing in the polls related to the number and size of campaign donations?

The 2020 Democratic primary season provides an unusually useful opportunity to study the relationship between fundraising and poll performance. With more than two dozen candidates vying for the nomination, there is a wealth of data to analyze. We draw on Federal Election Commission (FEC) data on campaign fundraising, along with daily aggregated polling data, to examine the timing and relationship between poll standing and campaign finance. While previous studies have modeled the relationship between financing and election outcomes via traditional methods such as time-series analysis, instrumental variables approaches, we introduce a technique previously used in health sciences research called joinpoint regression. We use this method to explore the relationship between donation data and poll standing, which provides a new perspective on measuring candidate campaign performance.

\section{METHODOLOGY}
This study applies joinpoint regression, a modeling technique previously used in health sciences research, to better understand the relationship between two key components of election campaigns: campaign finance and standing in the polls. 
Joinpoint regression can capture trend changes in a time series by modeling the outcome by a set of connected piece-wise linear segments. We use joinpoint regression modeling to explore the impact of one component on another, as well as their relationship to campaign performance.  In this section we discuss A. Data Collection, B. Data processing, and C. Joinpoint analysis. 

\subsection{Data collection}
Our study focused on two components: aggregated poll results and donation data collected from Federal Election Commision (FEC).

Different pollsters collect data with varying approaches such as phone calls and internet polls. They have distinct sampling sizes, sampling time, and potential biases in collecting the data. Appropriately aggregating the data from these polls helps to get a better estimation. The two well-known poll aggregators, FiveThirtyEight and RealClearPolitics, share aggregated values for the election. FiveThirtyEight also releases its rating for pollsters \cite{silver2014fivethirtyeight}. 

\begin{figure}[htbp]
\centerline{\includegraphics[width=90mm]{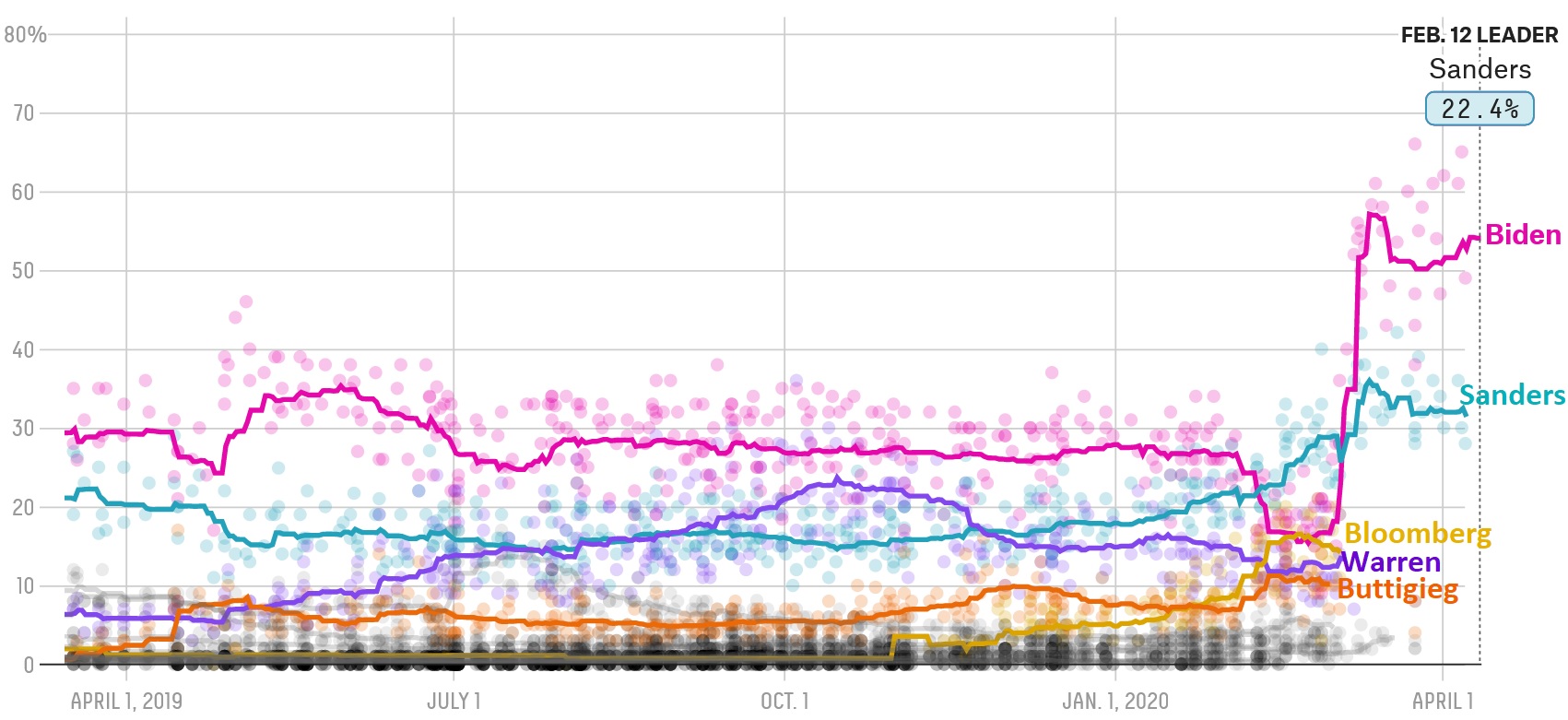}}
\caption{National average for each Democratic candidate in 2020 primary polls. Source: \cite{FiveThirthyEight_NA} }
\label{538}
\end{figure}

FiveThirtyEight collects poll data from different pollsters, rates them, finds their biases, and finally aggregates them to find the national average for different elections. The raw data, the rating, and biases of the pollsters are all available on FiveThirtyEight \cite{FiveThirthyEight_NA}. Fig.~\ref{538} from FiveThirtyEight shows the national average for each Democratic candidates in 2020 primary polls. 

Fig.~\ref{538}, shows the daily national average for each candidate but it is not easy to identify changes in the trend for each candidate nor see how significant events related to these changepoints. In the next section we use jointpoint regression to get more insight about this data. 

We collected the donation data provided by the FEC official website \cite{FEC:2020}. The raw donation data contains information from multiple aspects of the donation and donor including date, donor's name and zip-code. These information helps us distinguish the donors from each other, which provides a further possibility for counting and tracking each donor's donation behavior towards multiple candidates.

\subsection{Data processing}

Poll results represent estimated levels of support to each candidate on a given day.  Donation data measures performance of each campaign independently and its comparison with polls requires data processing. Here we briefly discuss how we processed the raw data collected from official FEC wesite.

Based on the information of each donation and its donor, we processed the data for each candidate and obtained four metrics: the daily number of donors, daily number of new donors, daily donation amount from each donor, and daily donation amount from each donor.

The daily donor number is the total number of donors who have donated to a candidate on a specific day; the daily donation amount is the total amount of donation a candidate has received in a specific day. The daily number of new donors is the number of donors who donated to the candidate for the first time on that day. 
Note that a donor who has never donated to candidate \emph{A} but has donated to candidate \emph{B} in the past, will still be distinguished as a new donor of candidate \emph{A} on the day of the donor's first donation to candidate \emph{A} no matter how many times the donor has donated to other candidates in the past; after the first donation to a specific candidate, a donor will not be distinguished as a new donor to this candidate. Correspondingly, the daily donation amount from new donors is defined as the total donation amount a candidate can receive from all the new donors on a specific day. These four metrics provide us the popularity and appeal of each candidate from the aspect of donation.

Moreover, we normalized the donation data by the total amount received over all candidates. This converts our measures into the daily proportion that went to each candidate. It is a two-step process: first, we find the sum of the total daily donations; second, the donations for each candidate are then divided by the sum of the total daily donations to represent the proportion.

\subsection{Joinpoint analysis}

A joinpoint regression model, also known as piecewise regression model or multi-phase regression, describes processes with multiple stages with variant trends. It can automatically detect the point that a change occurs as a ``joinpoint'', hence differentiating the stages into multiple segments. Each segment is modeled as a linear function with fluctuations which manifests its trend. Joinpoints correspond to the changes in trend and bridge the stages together. Traditional trend modeling, including linear regression and time series, are often overly smooth for solving real-world problems, Joinpoint regression, however, is able to handle these conditions better in some aspects because of its ability to segment the time series. Joinpoint regression can provide insight of the time component and its relationship to variables of interest in a way that ordinary least squares and simple regression models may miss. By clearly identifying the point at which trends change, it highlights specific points in time that warrant further analysis and these models can be overlapped to allow for insight of whether one variable may lag or stimulate change in another.

Joinpoint regression models are frequently implemented in health sciences. Example studies include using joinpoint regression in predicting the trends of disease rates \cite{kim2000permutation}, mortality rates of diseases \cite{fernandez2001recent}, and suicide rates \cite{hedegaard2020increase}. Joinpoint regression has been used outside of health sciences as shown by its application to detect trends between food pantry use and long-term unemployment \cite{shackman2015relation}. Joinpoint regression has also been applied to exploring relationships between disease rates and economic data to investigate the potential link or the existence of causality (economic variables affecting disease rates or vice versa) \cite{lubitz2014annual,regidor2014has}, where the intervention of economic crisis or financial alterations have been detected as the interpretable factor contributing to the fluctuation of health-related problems.

Joinpoint regression is a tool for trend comparison. Identifying the similarity of trends in relevant variables can be used to provide insight of one variable as a potential predictor of another. The change points, the points that represent major changes in the trend, can be a tool to explore the causality of  variables. The change points in the context of campaign performance can identify major events such as failure or success in a debate. 

Identifying important events that resulted in changes in the trend is another application of change points. Change points that occur closely after an event may provide insights into the effect of the event on the overall trends for a candidate.

Joinpoint analysis can also reveal relationship insight from the data. By checking the timing of increased campaign donations and changes in poll position, it can give insights as to which variable leads the other and under what conditions. With domain knowledge of the data, it is also possible to explore and find answers to additional  questions about the relationship between these variables. For example, can one be used to predict changes in the other? Does the relationship between polling and donations look the same at set milestones in the campaign or even across different candidates? And are there insightful trends in the data that can help distinguish the winning candidate from the others earlier on in the election process?

\section{Results}

In this project we used the \emph{R} package \emph{genlasso}  \cite{arnold2020package} for the joinpoint regression analysis. As a rule of thumb for this data we found having a degree of freedom equal to 12 for every three month can reveal the main dynamics of the data.

\subsection{Comparing different candidates}
\begin{figure}[htbp]
\centerline{\includegraphics[width=90mm]{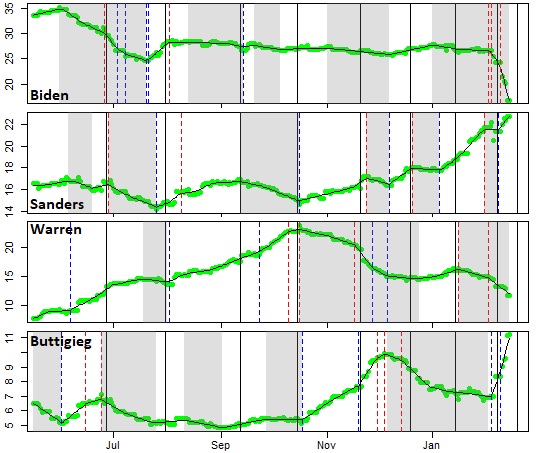}}
\caption{Joinpoint analysis for top Democratic candidates in 2020 primary polls. The fourth debate occurred on October 15, 2019. There is a noticeable impact on the change in trends for Bernard Sanders, Elizabeth Warren, and Pete Buttigieg following this event. The shift from rising to falling trends aligned on this date provides insight as to the success for these candidates in the debate.}
\label{fig_top_candid}
\end{figure}

We applied joinpoint regression to the aggregated polling data sets. Fig.~\ref{fig_top_candid} is an illustration of joinpoint regression. In the joinpoint regression plots the green dots are the average percentage of the daily national polls. The black trend line represents the smoothed regression of the joinpoint analysis. The black vertical lines represent debate dates. The dashed vertical lines represent the major joinpoints for each metric. Blue lines indicate the start point for an increase in the trend slope and red lines indicate the start point of a decrease in the trend slope. These trends are further identified by the grey background which highlights when a trend is falling while the white regions represent a rising slope or trend. 

Figure~\ref{fig_top_candid}  uses data specific to 2020 presedential candidates Joe Biden, Bernard Sanders, Elizabeth Warren, and Pete Buttigieg starting May 15, 2019 and going to February 15, 2020. These four candidates were selected as they were the top candidates in the polls for Democratic primary polls. 
 
We can use joinpoint analysis to identify and highlight event impacts for candidates. For example in Figure~\ref{fig_top_candid} we can observe that there is an alignment of trend shifts occurring in October. October 15, 2019 is when the fourth debate occurred and may be be the correlation of these aligned trend shifts. From the joinpoint plot it appears that Elizabeth Warren was impacted negatively by this event as her trend switched from a positive trend to negative. Pete Buttigieg an Bernard Sanders 

\begin{itemize}
    \item  Immediately after the fourth debate on October 15th, there are change points in the average polling result for Bernard Sanders, Elizabeth Warren, and Pete Buttigieg. Warren had experienced a great rising trend before this debate and she became one of the front-runners. Most of the other candidates attacked her she lost this debate. On the other hand,  Pete Buttigieg and Bernard Sanders were among the winners of this debate \cite{fourth_debate}. 
    \item For longer than a month Warren experienced a falling trend after the fourth debate but Bernard Sanders and Pete Buttigieg experienced a rising trend. It looks what Warren lost was a gain for the other two candidates.
    \item Pete Buttigieg was one of the winners in the fifth debate\cite{fifth_debate} and we can observe its effect on the trends after the debate.  
\end{itemize}

\subsection{Different metrics for one candidate}
\begin{figure}[htbp]
\centerline{\includegraphics[width=90mm]{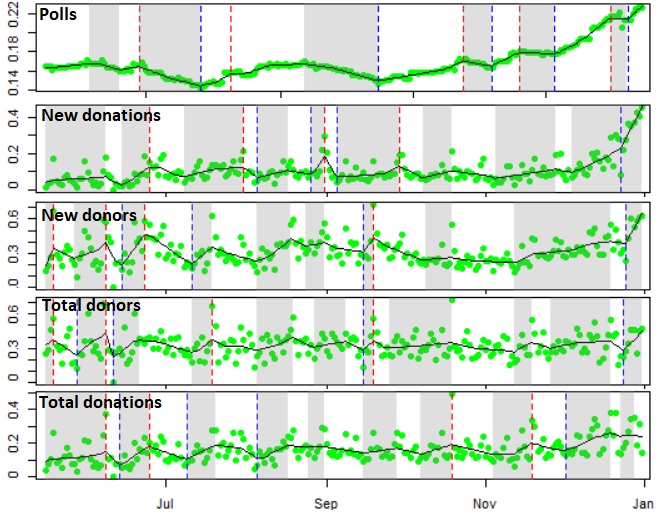}}
\caption{Joinpoint regression on different metrics for candidate Sanders. We can observe, most of the times, a change point in the polling data occurs very close to a change point in at least one of the donation data and the steep slopes in the polling data occurs when there are similar steep slopes in the donation data. }
\label{fig_sanders}
\end{figure}

 We applied joinpoint regression to both the aggregated polling and the donation data sets in Fig.~\ref{fig_sanders}. This figure  uses data specific to the  Bernard Sanders who was one of the top candidate in the polls at that time. The green dots are daily poll aggregations, new donations, new donors, total donors and total donation  . The black line through them is the trend for a specific variable. 

We can observe the followings in Fig.~\ref{fig_sanders},

\begin{itemize}
    \item The falling and rising trends across different donation metrics are similar. This fact implies their dynamic is very similar.
    \item Most of the times, a change point in the polling data occurs very close to a change point in at least one of the donation data. 
    \item Steep slopes in the polling data occurs when there are similar steep slopes in the donation data. As a result, steep slopes in one metric can indicate similar trend in the other. 
\end{itemize}
These observation implies that polling and donation data are alternative measure of campaign performance. When there is uncertainty about the accuracy of one, the other improves the interpretation of that metric. 

\begin{figure}[htbp]
\centerline{\includegraphics[width=90mm]{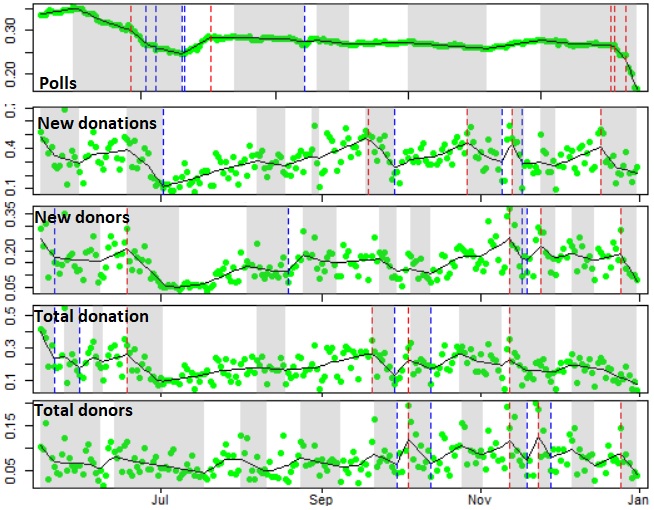}}
\caption{Joinpoint regression on different metrics for Joe Biden }
\label{fig_biden}
\end{figure}

Fig.~\ref{fig_biden} similarly uses data specific to the candidate Joe Biden for the same period.

\subsection{Identifying events}
\begin{figure}[htbp]
\centerline{\includegraphics[width=90mm]{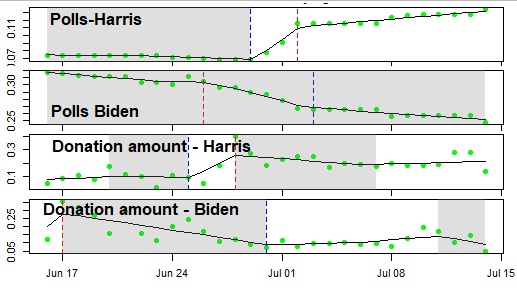}}
\caption{Change of trends for candidates Joe Biden and Kamala Harris close to the first debate. 
The second night of this debate occurred on June 27, 2019. The debate between candidates Joe Biden and Kamala Harris was one of the most important parts of this debate and there was a noticeable impact on the change in trends for these two candidates. They continued this fight even after the debate. The two change points in their polling data in a two week period may highlight this event.}
\label{biden_harrs}
\end{figure}

The change points can help identifying the events that caused change of the trends. For example, looking at poll results for Kamala Harris and Joe Biden in Fig.~\ref{biden_harrs} we can find two close change points in late June and early July.

\begin{figure}[htbp]
\centerline{\includegraphics[width=90mm]{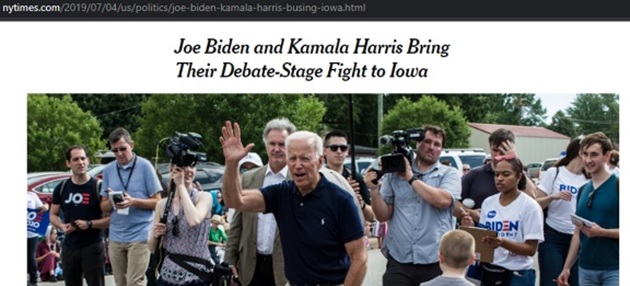}}
\caption{New York Times headline on July 4th. Source: \cite{NYT_Biden_H}}
\label{NYD}
\end{figure}

Checking the headlines from New York Times in that period shown in Fig.~\ref{NYD} we can find how her fight with Joe Biden in the first debate helped her to become more successful. On the other hand, in the same period of time, Joe Biden was less successful comparing to the trend before the debate.

\subsection{Dynamics in different metrics}

\begin{figure}[htbp]
\centerline{\includegraphics[width=90mm]{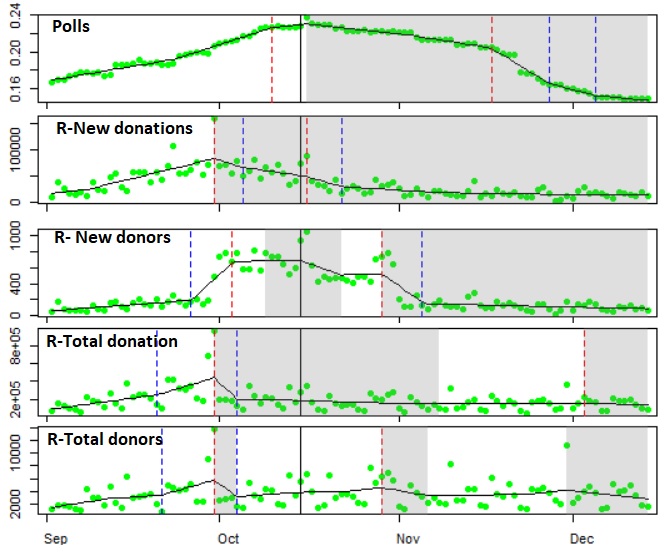}}
\caption{Joinpoint regression on polls and raw donation data for  Elizabeth Warren. One change point in the late September and one in mid-October can be observed. The first one might be associated with the deadline for the requirements of the fourth debate and the debate on October 15th.  }
\label{fig_warren}
\end{figure}

So far we have focused on how these metrics compare across candidates. Another application of this analysis is finding the different dynamics across metrics for one candidate. In this case, we don't need to normalize the data based on their daily summations.  We can apply joinpoint regression on the raw data in this case. In Fig.~\ref{fig_warren}, we can observe the result of joinpoint regression on the raw donation for  Elizabeth Warren.

In the polling data shown in Fig.~\ref{fig_warren},  Elizabeth Warren had a rising trend before mid-October and a falling trend afterward. Mid-October is the time of the fourth debate and most of the other candidates attacked her on the debate since she had became one of the front-runners. 

\begin{figure}[htbp]
\centerline{\includegraphics[width=90mm]{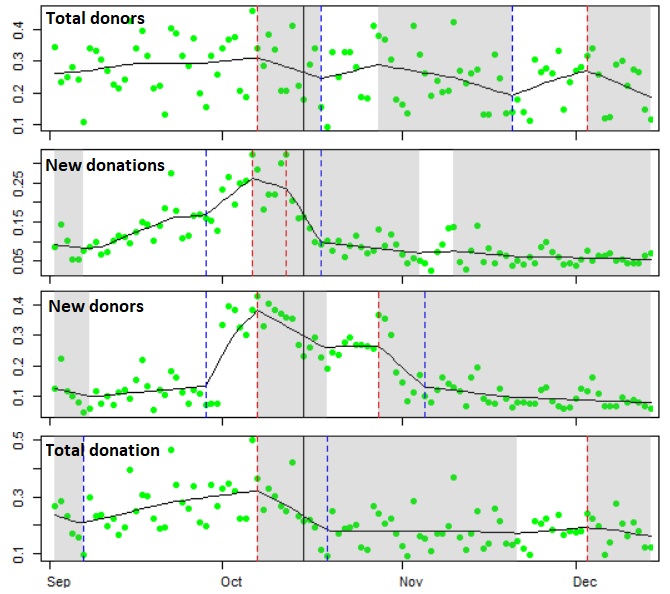}}
\caption{Joinpoint regression on different normalized metrics for  Elizabeth Warren. Comparing these plots with Fig.~\ref{fig_warren} we can observe the change points at the end of the September shifted to early October. This can be result of the normalization and increased donation for all other candidates.   }
\label{fig_warren2}
\end{figure}

If we look at the donation data in Fig.~\ref{fig_warren}, all the metrics had a change point in early October. October 1st was the deadline for satisfying the fourth debate requirements. It is not surprising that her campaign asked her supporters to donate before this deadline. Other candidates had the same deadline and so if we normalized the donation data we expect the trends to be different. In Fig.~\ref{fig_warren2} the normalized donation data is shown. As we can see, the change points are shifted to the right. 

Comparing Fig.~\ref{fig_warren2} and Fig.~\ref{fig_warren}, we observe the blue change points occur after the fourth debate in the normalized donation data but not the raw data. It explains these change points are mainly resulted from increased support for other candidates which affects the denominator in the normalizing process.

\section{DISCUSSION}

We can observe the sequence of joinpoints are usually similar for each of the metrics graphed meaning the trends of rises and falls in polling are also reflected in donation measures and vice versa. Donation metrics lead the polls at times as shown in late June however this does not remain consistent; sometimes we observe a lag in donation metrics as shown in early August. 

The first Democratic presidential debate occurred on June 26th and June 27th, 2019. Looking at that time period we can observe that Joe Biden had negative trends in polling and donation data shortly after the debate. This indicates the debate may not have been helpful for his campaign as his position in the polls continued to decline. This debate presented an opportunity for lesser known candidates to have an introduction to the public and it may represent their success at exposure more than a mistake or failure of Joe Biden’s. This is consistent with previously published literature citing primary debates as the time when voters are seeking information to learn about new candidates and clarify their position \cite{mckinney2013presidential}.

Based on data from a single candidate it’s unclear whether one variable can be used as a predictor for the other. It is our intention to further this analysis exploring overarching trends by comparing data across a wider range of time and multiple candidates.  Additionally, joinpoint regression analysis is susceptible to apriori setting of the number of joinpoints that exist over the time period. This means that incorrectly setting the number of joinpoints may cause a misrepresentation of the data or cause one to miss key changes in the trends of the variables. This is a topic that would need to be explored further in order to understand the impact of the number of joinpoints on this type of data set.

\section{CONCLUSION}
To our knowledge, this study is the first time that the joinpoint regression model has been implemented in the political domain or in predicting aggregated election polls of presidential candidates predicted by donations received from individual supporters. Distinctive from traditional linear or time series modeling, joinpoint regression modeling provides a novel and valuable perspective of investigating the relationship between polls of candidates and their fundraising results. 

This study accomplishes two key advances: it builds a foundation for future exploration using joinpoint regression by crossing joinpoint analysis into the political domain and it demonstrates the value of joinpoint regression in political analysis. The models illustrated in this study represent some of the insights that can be used within the political domain to bolster campaign analysis presently. In future studies and analysis we plan to explore the consistency of joinpoint trends across wider ranging variables to assess overarching trends.

For the future work we will have more analysis on the consistency of the change points and whether one metric leads the other one and what is the main reasoning for it. Also, we would have some theoretical discussions such as the optimal degree of freedom for this analysis.

\bibliographystyle{IEEEtran}
\bibliography{refs}

\end{document}